# Regularization by ε-metric

V.D. Ivashchuk

VNIIMS & RUDN University,

Moscow, Russia

*Abstract.* The regularization of propagators by means of a complex metric is considered. [*)]

**Introduction.** In quantum field theory and gravity, one often has to deal with pseudo-Euclidean-type peculiarities. These peculiarities [singularities] are regularized by introduction of complex parameters. For example, the covariant-regularized propagator of a free massive scalar field in the momentum representation has the form [1]

$$(\overline{\varphi\varphi})(p, \varepsilon) = i/(p^2 - m^2 + i\varepsilon), \qquad (1)$$

where $\varepsilon > 0$. Regularization (1) has two significant drawbacks, namely that it does not guarantee:

1) convergence of Feynman integrals at ε > 0, even if the corresponding integrals of the Euclidean theory converge;
2) the regularity of propagators in the coordinate representation, even if the specified regularity takes place in the Euclidean case.

(Here and below all integrals are Lebesgue integrals [2]. Relevant counter-examples will be given in section 2.)

This paper proposes an alternative regularization scheme free of the above drawbacks. In this scheme the propagators are regularized by using the "ε-metric" (see section 3):

$$\left(\eta^{\varepsilon}_{\alpha\beta}\right) = \begin{pmatrix} e^{-i\varepsilon} & 0 & \ldots & \ldots & 0 \\ 0 & -1 & \ldots & \ldots & 0 \\ \ldots & \ldots & \ddots & \ldots & \vdots \\ \ldots & \ldots & \ldots & -1 & 0 \\ 0 & \ldots & \ldots & 0 & -1 \end{pmatrix}, \quad 0 < \varepsilon < 2\pi. \qquad (2)$$

________________________________________________________________

In this regularization, for example, instead of (1), the following relation is obtained (see section 3):

$$(\underline{\varphi\varphi})_\varepsilon(p) = i\, e^{i\varepsilon/2} / (p_\varepsilon^2 - m^2), \qquad (3)$$

where $p_\varepsilon^2 = e^{i\varepsilon} p_0^2 - \mathbf{p}^2$.

If $m^2 > 0$, then both regular generalized functions [distributions] of moderate growth (1) and (3) have a limit at $\varepsilon \to +0$ in $S'(R^D)$ $(D \geq 2)$ and both limits coincide. (The next article of the author will be devoted to the limiting transition $\varepsilon \to +0$.)

"Metric" (2) naturally implements the idea of Euclidean reversal [Wick rotation]:

A) $\varepsilon = \pi$ corresponds to the Euclidean space (E);

B) $\varepsilon = +0$ – to Minkowski space [for D =4] with the correct direction of the arrow [of time] - "our world";

C) $\varepsilon = 2\pi - 0$ – to Minkowski space [for D =4] with the wrong direction of the arrow [of time] $(M_\leftarrow)$ – "antiworld" (see Figure below).

A direct generalization of (2) is the so-called "$a$ - metric":

$$(\eta_{\alpha\beta}) = \begin{pmatrix} a^{-1} & 0 & \cdots & \cdots & 0 \\ 0 & -1 & \cdots & \cdots & 0 \\ \cdots & \cdots & \ddots & \cdots & \vdots \\ \cdots & \cdots & \cdots & -1 & 0 \\ 0 & \cdots & \cdots & 0 & -1 \end{pmatrix}, \qquad (4)$$

where $a \in C \setminus [0, +\infty)$. In fact, our regularization prescription (see section 3) consists in an analytical extension of the parameter $a$ from the Euclidean domain. At the same time, figuratively speaking, "our world" is "non-analytical in metric" (i.e. in $a$), "we are on the upper bank of the section (see figure), antiworld - on the lower."

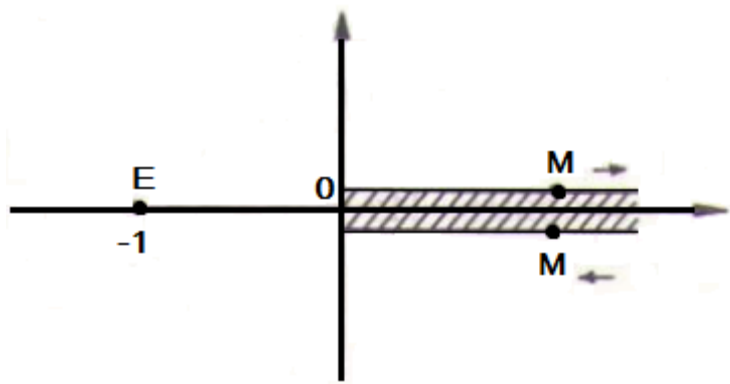

[Figure. The domain of complex plane for $a$-parameter.]

The plan of the presentation of the proposed scheme of the regularization is following: in section 3 the prescription of the regularization is formulated; in section 4 an α-representation for the regularized propagator (3) is introduced; on an example of a correlator of a scalar field of a very general kind it is shown how from the Euclidean correlator one can obtain ε-correlator in the coordinate, momentum and α-representations; the concept of admissible (proper) ε-correlator is introduced.

**2. Covariant regularization and its drawbacks.** As it was mentioned in section 1, the covariant introduction of the ε-term (1) does not guarantee the existence of a Feynman integral, even if the corresponding integral of the Euclidean theory exists. Let us present a simple example:

$$\int d^3k \frac{1}{(k^2-m^2+i\varepsilon)((k+q)^2-m^2+i\varepsilon)} \tag{5}$$

$m^2 > 0, \varepsilon > 0$.

The corresponding integral of Euclidean theory:

$$\int d^3k \frac{1}{(k_E^2 + m^2)((k+q)_E^2 + m^2)}$$

for $m^2 > 0$ and $\forall\, q \in R^3$ does exist. Here $k_E^2 = k_0^2 + \mathbf{k}^2$.

**Proposition 1.** *The Lebesgue integral* (5) *for all values* $m^2 > 0, \varepsilon > 0$ и $q \in R^3$ *does not exist*.

**Proof.** Let us suppose the opposite – the integral (5) exists. It is known [2], that the integrability of a complex-valued function $f$ on $R^D$ implies the integrability of the function $|f|$ on $R^D$, and hence the integrability on any measurable subset. Consider the domain $V = \{k | \sqrt{\mathbf{k}^2 + m^2} \leq k_0 \leq \sqrt{\mathbf{k}^2 + 2m^2}\}$. $V$ is open in $R^D$, hence it is measurable. From supposing about the existence of (5) the existence of the integral follows:

$$\int_V d^3k \frac{1}{|k^2-m^2+i\varepsilon||(k+q)^2- m^2-i\varepsilon|}. \tag{6}$$

For $k \in V$ we have:

$$|k^2 - m^2 + i\varepsilon| \leq |k^2 - m^2| + \varepsilon$$

$$|(k+q)^2 - m^2 - i\varepsilon| \leq \left||k^2 - m^2| + 2|kq| + |q|^2 + \varepsilon\right|$$

$$\leq m^2 + |q^2| + \varepsilon + 2|q_0|\sqrt{\mathbf{k}^2 + 2m^2} + 2|\mathbf{k}||q|.$$

From this we have for $k \in V$

$$\frac{1}{m^2 + |q^2| + \varepsilon + 2|q_0|\sqrt{\mathbf{k}^2 + 2m^2} + 2|\mathbf{k}||q|}$$
$$\leq \frac{m^2 + \varepsilon}{|k^2 - m^2 + i\varepsilon||(k+q)^2 - m^2 + i\varepsilon|}. \tag{7}$$

In the left hand side of (7) the function of variable $k$ is continuous, so it is measurable. Then from (7) and the existence of the integral (6) the existence of the integral follows

$$\int_V d^3k \frac{1}{m^2 + |q^2| + \varepsilon + 2|q_0|\sqrt{\mathbf{k}^2 + 2m^2} + 2|\mathbf{k}||q|}. \tag{8}$$

According to Fubini's theorem [2], there exists a repeated integral equal to [8]:

$$\int d^2\mathbf{k} \int_{\sqrt{\mathbf{k}^2+m^2}}^{\sqrt{\mathbf{k}^2+2m^2}} dk_0 \frac{1}{m^2 + |q^2| + \varepsilon + 2|q_0|\sqrt{\mathbf{k}^2+2m^2} + 2|\mathbf{k}||q|} =$$
$$\int d^2\mathbf{k} \frac{\sqrt{\mathbf{k}^2+2m^2} - \sqrt{\mathbf{k}^2+m^2}}{m^2 + |q^2| + \varepsilon + 2|q_0|\sqrt{\mathbf{k}^2+2m^2} + 2|\mathbf{k}||q|} =$$
$$m^2 \int d^2\mathbf{k} \frac{1}{(\sqrt{\mathbf{k}^2+2m^2} + \sqrt{\mathbf{k}^2+m^2})(m^2 + |q^2| + \varepsilon + 2|q_0|\sqrt{\mathbf{k}^2+2m^2} + 2|\mathbf{k}||q|)} \tag{9}$$

However, the integral (9) obviously diverges for all $q \in R^3, m^2, \varepsilon > 0$. The obtained contradiction proves the proposition.

Nevertheless, there exist ways of "computing" non-existing integrals of the form (5), which lead to meaningful relations and in the limit $\varepsilon \to +0$ give well-defined generalized functions [distributions].

The first such "method" is based on the α-representation [1] of regularized propagator (1):

$$\frac{i}{p^2 - m^2 + i\varepsilon} = \int_0^\infty d\alpha e^{i(p^2 - m^2 + i\varepsilon)\alpha}, \qquad \varepsilon > 0. \tag{10}$$

In integrals of the form (5) propagators are written in the form (10) and then the order of integration in momenta and α is changed. The integral in momenta is understood as a limit

$$\lim_{\delta \to +0} \prod_i d^D k_i e^{-\delta \sum_i (k_i)_E^2} \{\ldots\}.$$

In the case of integral (5), as it is not difficult to check, a convergent "repeated" integral may be obtained for all $m^2, \varepsilon > 0$ and $q \in R^3$.

The second way in application to the integral (5) is as follows: we "cut" the integral (5), i.e. we integrate in (5) by ball $V_R = \{k | k_E^2 \leq R^2\}$. The "cut" integral does exist. If $R \to \infty$, then the limit

$$\lim_{R \to \infty} \int_{V_R} d^3k \frac{1}{(k^2 - m^2 + i\varepsilon)((k+q)^2 - m^2 + i\varepsilon)} \tag{11}$$

exists for all $m^2, \varepsilon > 0$ и $q \in R^3$ and coincides with the integral obtained by the first method. This does not contradict Proposition 1, since the existence of the limit (11) does not imply the existence of Lebesgue integral (5).

Now we give a counterexample confirming that covariant regularization (1) does not guarantee the regularity of the propagator in the coordinate representation, even if the specified regularity takes place in the Euclidean case.

In fact, it is valid in $S'(R^4)$

$$F^{-1}\left(\frac{1}{m^2 + p_E^2}\right)(x) = \frac{mK_1\left(m(x_E^2)^{1/2}\right)}{4\pi^2 (x_E^2)^{1/2}}, \tag{12}$$

where $m > 0$; $K_1(z)$ is the MacDonald function. Here and below $F$ и $F^{-1}$ are direct and inverse Fourier transforms of $S'(R^D)$, which are normalized on $S(R^D)$ by relations:

$$F(\varphi)(p) = \int d^D x e^{ipx} \varphi(x),$$

$$F^{-1}(\varphi)(p) = \int \frac{d^D p}{(2\pi)^D} e^{-ipx} \varphi(p),$$

where $\varphi \in S(R^D)$.

Function (12) is regular. We call by regular functions in $S'(R^D)$ those generalized functions which are generated by the "slowly integrable" functions on $R^D$. Complex valued function $f$ on $R^D$ is "slowly integrable" if for some $N$ the function $f(1 + x_E^2)^{-N}$ is integrable on $R^D$. We identify slowly integrable functions with generated by them [generalized] functions from $S'(R^D)$.

Meanwhile, it can be shown that in $S'(R^D)$ the following relation is valid

$$F^{-1}\left(\frac{1}{p^2 - m^2 + i\varepsilon}\right)(x)$$
$$= \frac{1}{4\pi^2(-x^2 + i0)}$$
$$+ (m^2 - i\varepsilon)^{1/2} \frac{\bar{K}_1\left((m^2 - i\varepsilon)^{1/2}(-x^2 + i0)^{1/2}\right)}{4\pi^2(-x^2 + i0)^{1/2}}. \tag{13}$$

where $\bar{K}_1(z) = K_1(z) - 1/z$, and

$$F^{-1}\left(\frac{1}{p^2 - m^2 + i\varepsilon}\right) \bar{\in} \operatorname{reg} S'(R^4) \ .$$

[I.e. we get non-regular generalized function.]

**3. The prescription of regularization by ε-metric.** As it was defined in section 1, ε-metric is a diagonal $D \times D$ matrix (2):

$$(\eta^\varepsilon_{\alpha\beta}) = \operatorname{diag}(e^{-i\varepsilon}, -1, \ldots, -1),$$

where $D$ is the dimension of the space-time, $0 < \varepsilon < 2\pi$. Inverse of (2) matrix is

$$(\eta^{\alpha\beta}_\varepsilon) = (\eta^\varepsilon_{\alpha\beta})^{-1} = \operatorname{diag}(e^{i\varepsilon}, -1, \ldots, -1). \tag{14}$$

The prescription of the regularization is that one should formally deal with ε-metric as with pseudo-Euclidean [real] one.

We put $x^\alpha$ и $p_\beta$ (i.e. coordinates with an upper index, and the components of the momentum with a lower index) to be real. Then

$$(xy)_\varepsilon \equiv \eta^\varepsilon_{\alpha\beta} x^\alpha x^\beta = e^{-i\varepsilon} x^0 y^0 - \mathbf{x} \cdot \mathbf{y};$$

$$x^2_\varepsilon = (xx)_\varepsilon = e^{-i\varepsilon}(x^0)^2 - \mathbf{x}^2; \tag{15}$$

$$(pq)_\varepsilon \equiv \eta^{\alpha\beta}_\varepsilon p_\alpha q_\beta = e^{i\varepsilon} p_0 q_0 - \mathbf{p} \cdot \mathbf{q};$$

$$p^2_\varepsilon \equiv (pp)_\varepsilon = e^{i\varepsilon} p^2_0 - \mathbf{p}^2;$$

$$px \equiv p_\alpha x^\alpha,$$

where $x$ and $y$ are [sets of] coordinates, and $p$ and $q$ are [sets of] momenta.

Volume forms by analogy with the pseudo-Euclidean case are replaced by

$$(d^D x)^2_\varepsilon = \sqrt{{}_2(-1)^{D-1} \det(\eta^\varepsilon_{\alpha\beta})}\, d^D x = e^{-i\varepsilon/2}\, d^D x, \tag{16}$$

$$(d^D p)^2_\varepsilon = \sqrt{{}_1(-1)^{D-1} \det(\eta^{\alpha\beta}_\varepsilon)}\, d^D x = e^{i\varepsilon/2}\, d^D p.$$

Here and in what follows

$$\sqrt{{}_2 Z} = -\sqrt{{}_1 Z}$$

are branches of a square root with a cut $[0, +\infty)$ normalized by the condition:

$$\sqrt{{}_1}(1+i0) = 1.$$

Regularization is carried out by replacing the continual [functional] measure

$$D\varphi \, exp(iS(\varphi)) \to D\varphi \, exp(iS_\varepsilon(\varphi)), \qquad (17)$$

where $S_\varepsilon(\varphi)$ has "$\varepsilon$-covariant" form and

$$S_{\varepsilon=+0}(\varphi) = S(\varphi).$$

We explain the construction of $S_\varepsilon(\varphi)$ by specific examples.

**Example 1. Free scalar field.**

Here

$$S_\varepsilon(\varphi) = \frac{1}{2}\int (d^D x)_\varepsilon ((\partial\varphi)_\varepsilon^2 - m^2\varphi^2), \qquad (18)$$

where

$$(\partial\varphi)_\varepsilon^2 = \eta_\varepsilon^{\alpha\beta}(\partial_\alpha\varphi)\partial_\beta\varphi = e^{i\varepsilon}(\partial_0\varphi)^2 - (\boldsymbol{\partial}\varphi)^2.$$

The measure (17) with the $\varepsilon$-action (18) generates the $\varepsilon$-correlator

$$\langle\varphi(x)\varphi(0)\rangle_\varepsilon = \underline{(\varphi(x)\varphi(0))}_\varepsilon = F^{-1}\left(\underline{(\varphi\varphi)}_\varepsilon(p)\right)(x) \qquad (19)$$

where $\underline{(\varphi\varphi)}_\varepsilon(p) = i\, e^{i\varepsilon/2}/(p_\varepsilon^2 - m^2)$ is $\varepsilon$-correlator in the momentum representation (3). We note that in the Euclidean case ($\varepsilon = \pi$) we obtain

$$\underline{(\varphi\varphi)}_{\varepsilon=\pi}(p) = \frac{1}{p_E^2 + m^2} = \underline{(\varphi\varphi)}_E(p)$$

and

$$D\varphi \, exp(iS_{\varepsilon=\pi}(\varphi)) = D\varphi \, exp(-S_E(\varphi)),$$

where

$$S_E(\varphi) = \frac{1}{2}\int d^D x ((\partial\varphi)_E^2 + m^2\varphi^2).$$

That is, ε-metric describes the Euclidean case for $\varepsilon = \pi$.

**Example 2. Free Fermi field (D=4).**

In this case

$$S_\varepsilon(\psi) = \int (d^4x)_\varepsilon \, \bar\psi(i\hat\partial_\varepsilon - m)\psi \tag{20}$$

and

$$\left(\psi\bar\psi\right)_\varepsilon(p) = \frac{i(m + \hat p_\varepsilon)}{p_\varepsilon^2 - m^2}. \tag{21}$$

Here

$$\hat p_\varepsilon = p_\alpha \gamma_\varepsilon^\alpha; \quad \hat\partial_\varepsilon = \gamma_\varepsilon^\alpha \partial_\alpha; \tag{22}$$

$$\gamma_\varepsilon^0 = e^{i\varepsilon/2}\, \gamma^0; \quad \gamma_\varepsilon^i = \gamma^i, \quad i = 1,2,3,$$

where modified $\gamma_\varepsilon$-matrices satisfy

$$\gamma_\varepsilon^\alpha \gamma_\varepsilon^\beta + \gamma_\varepsilon^\beta \gamma_\varepsilon^\alpha = 2\,\eta_\varepsilon^{\alpha\beta}. \tag{23}$$

In fact, there is an implicit ε-bein in (20) - (23)

$$(e_\alpha^{a\varepsilon}) = \mathrm{diag}(e^{-i\varepsilon/2}, 1, 1, 1);$$

$$\eta_{\alpha\beta}^\varepsilon = \eta_{ab} e_\alpha^{a\varepsilon} e_\beta^{b\varepsilon}.$$

**Example 3. Yang-Mills field.**

Here

$$A_\alpha = A_\alpha^a t_a \in \mathfrak{g}\text{ - semisimple Lie algebra,}$$

$$S_\varepsilon(A) = -\frac{1}{4}\langle F_{\alpha\beta}(A), F_\varepsilon^{\alpha\beta}(A)\rangle, \tag{24}$$

where

$$F_{\alpha\beta}(A) = \partial_\alpha A_\beta - \partial_\beta A_\alpha + [A_\alpha, A_\beta];$$

$$F_\varepsilon^{\alpha\beta}(A) = \eta_\varepsilon^{\alpha\alpha'} \cdot \eta_\varepsilon^{\beta\beta'} F_{\alpha'\beta'}(A),$$

$\langle .,.\rangle$ is Killing form.

The "free" ε-correlator in $\tilde\alpha$-gauge (in momentum representation) may be obtained from (24) in a standard way [4]:

$$(A^a_\alpha A^b_\beta)_\varepsilon(p) = \frac{e^{i\varepsilon/2}g^{ab}}{ip_\varepsilon^2}\left(\eta^{\alpha\beta}_\varepsilon - \frac{p_\alpha p_\beta}{p_\varepsilon^2}(1-\tilde{\alpha})\right), \qquad (25)$$

where

$$(g^{ab}) = (g_{ab})^{-1}; \quad g_{ab} = \langle t_a, t_b \rangle.$$

We also present a formula that determines the dimensional regularization in the "ε - metric". The following relation may be verified by a direct calculation:

$$\int \frac{(d^D k)_\varepsilon}{i(C-k_\varepsilon^2)^\alpha} = \frac{\pi^{D/2}}{C^{\alpha-D/2}}\frac{\Gamma(\alpha-D/2)}{\Gamma(\alpha)}, \qquad (26)$$

for $\alpha > D/2$ and

$$-\pi < \arg C < \varepsilon, \quad \text{if } 0 < \varepsilon \leq \pi;$$

$$\varepsilon - 2\pi < \arg C < \pi, \quad \text{if } \pi \leq \varepsilon < 2\pi;$$

When $\varepsilon = \pi$ we obtain from (24) the Euclidean relation:

$$\int \frac{d^D k}{(C+k_E^2)^\alpha} = \frac{\pi^{D/2}}{C^{\alpha-D/2}}\frac{\Gamma(\alpha-D/2)}{\Gamma(\alpha)},$$

where $\alpha > D/2$ and $|\arg C| < \pi$. The formula (26) may be taken as a starting point for determining the dimensional regularization in the ε-metric.

*Majorizing inequalities.* A remarkable property of the regularization introduced by us is the fact that the behavior of ε-correlators at high momenta is equivalent to the behavior of the corresponding Euclidean correlators. This follows from the following chain of inequalities:

$$(\sin(\varepsilon/2))(k_E^2 + m^2) \leq |k_\varepsilon^2 - m^2| \leq k_E^2 + m^2, \qquad (27)$$

which are valid for all $\varepsilon\ [0 < \varepsilon < 2\pi], m$ и $k$. The right hand side inequality in (27) is trivial, the left hand side inequality is deduced as very simple one:

$$|k_\varepsilon^2 - m^2| = \left|e^{-i\varepsilon/2}(k_\varepsilon^2 - m^2)\right| \geq \left|\text{Im}\left(e^{-i\varepsilon/2}(k_\varepsilon^2 - m^2)\right)\right|$$

$$= (\sin(\varepsilon/2))(k_E^2 + m^2).$$

By virtue of (27), the Feynman integral composed of ε-correlators (3) exists for $0 < \varepsilon < 2\pi$, if it exists in the Euclidean case. This statement can be extended to a wider class of ε-correlators.

**Remark.** ε-metric is a special case of $a$-metric (4):

$$\eta^\varepsilon_{\alpha\beta} = \eta_{\alpha\beta}(e^{i\varepsilon}).$$

All relations given for ε-metric are easily generalized to the case of $a$-metric. For example, the relation (3) is replaced by the following one:

$$(\underline{\varphi\varphi})(p,a) = \frac{(-a)^{1/2}}{m^2-p^2(a)} \tag{28}$$

where $a \in C\setminus[0,+\infty)$ and $p^2(a) = \eta^{\alpha\beta}(a)\, p_\alpha p_\beta = a\, p_0^2 - \mathbf{p}^2$,

[here $(\eta^{\alpha\beta}(a)) = (\eta_{\alpha\beta}(a))^{-1}$ ].

For $a \in C\setminus[0,+\infty)$ the following inequalities are valid:

$$\left(\sin\left(\frac{\arg_1 a}{2}\right)\right)\frac{|a|}{1+|a|}(k_E^2 + m^2) \le |k^2(a) - m^2| \\ \le (1+|a|)(k_E^2 + m^2), \tag{29}$$

where $\arg_1(.)$ is a branch of the argument in $C\setminus[0,+\infty)$,

$$\arg_1(1 + i0) = 0.$$

**4. α-representation. Proper ε-correlator.**

ε-correlator (3) admits an α-representation. For $m^2 > 0$, $0 < \varepsilon < 2\pi$ and all $p \in R^D$ the following relation is valid:

$$\frac{ie^{i\varepsilon/2}}{p_\varepsilon^2 - m^2} = \int_0^\infty d\alpha \exp\left(i\,\alpha\, e^{-i\varepsilon/2}\,(p_\varepsilon^2 - m^2)\right). \tag{30}$$

The convergence of the integral in (30) follows from the identity

$$\left|\exp\left(ie^{-i\varepsilon/2}\,\alpha(p_\varepsilon^2 - m^2)\right)\right| = \exp\left(-\alpha\,\sin(\varepsilon/2)(p_E^2 + m^2)\right). \tag{31}$$

**Definition 1.** *Given an* ε-correlator $(\underline{\Phi\Phi})_\varepsilon(p)$, $\varepsilon \in (0,2\pi)$. *We call it proper, if for all* $\varepsilon \in (0,2\pi)$, *it can be presented as*

$$(\underline{\Phi\Phi})_\varepsilon(p) = P(p,\varepsilon)\int_0^\infty d\alpha\, f(i\alpha\, e^{-i\varepsilon/2})\exp\left(i\alpha\, e^{-i\varepsilon/2}(p_\varepsilon^2 - m^2)\right), \tag{32}$$

where

$1^0$. A) $f(\alpha)$ *is holomorphic in the region* $\{\operatorname{Re}\alpha > 0\}$ *and continuous in* $\{\operatorname{Re}\alpha \ge 0\}\setminus\{0\}$;

B) $f(\alpha) = O(\alpha^T)$ *for* $\alpha \to \infty$, $\operatorname{Re}\alpha \ge 0$;

C) *for some* $s > -1$ *and all* $\delta > 0$ *the relations are valid*

$$f(\alpha) = O(\alpha^{s-\delta}),$$
$$\alpha^{s+\delta} = O(f(\alpha)),$$

when $\alpha \to 0$, $\operatorname{Re}\alpha \geq 0$;

$2^0$. $m^2 > 0$;

$3^0$. *There exists a holomorphic function $F(z)$ defined in $\mathbb{C}\setminus(-\infty, 0]$, which for $\operatorname{Re} z > 0$ is equal to*

$$F(z) = \int_0^\infty d\alpha f(\alpha) e^{-\alpha z} ; \qquad (33)$$

$4^0$. *$P(p, \varepsilon)$ is a polynomial in $p_\alpha$ with coefficients (possibly matrix-valued) smoothly depending on $\varepsilon$, and $P(p, 0)$ is covariant.*

Note, that due to $1^0$, $2^0$ and (31) the integral in (32) exists for all p and $\varepsilon \in (0, 2\pi)$, free field $\varepsilon$-correlators (3) and (21) satisfy the conditions of the Definition 1.

Definition 1 is natural. This can be seen in the example of a scalar field. If the Euclidean correlator is regular in coordinate and momentum representations and has the form

$$\underline{(\varphi(x)\varphi(0))}_E = \tilde{D}(x_E^2), \qquad (34)$$

$$\underline{(\varphi\,\varphi)}_E(p) = D(p_E^2),$$

where

$$D(p_E^2) = \int_0^\infty d\alpha f(\alpha) \exp\left(-\alpha(p_E^2 + m^2)\right), \qquad (35)$$

$m^2 > 0$, then under some restrictions on $D(z), \tilde{D}(z), f(\alpha)$ on the $a$-correlator $(\varphi(x)\varphi(0))\,(a)$ it can be obtained that

$$\underline{(\varphi(x)\varphi(0))}\,(a) = = \tilde{D}(-x^2(a)), \qquad (36)$$

$$\underline{(\varphi\,\varphi)}(p, a) = (-a)^{1/2} D(-p^2(a))$$

and

$$(-a)^{1/2} D(-p^2(a)) = \int_0^\infty d\alpha f(\alpha(-a)^{-1/2}) \exp\left(-\alpha(-a)^{-1/2}(m^2 - p^2(a))\right) \quad (37)$$

for $a \in \mathbb{C}\setminus[0, +\infty)$.

For $\varepsilon$-metric ($a = e^{i\varepsilon}$) we have

$$\underline{(\varphi(x)\varphi(0))}_\varepsilon = \tilde{D}(-x_\varepsilon^2), \qquad (38)$$

$$(\underline{\varphi\,\varphi})_\varepsilon(p) = (e^{i\varepsilon/2}/i)\ D(-p_\varepsilon^2)$$

and

$$\frac{e^{i\varepsilon/2}}{i}D(-p_\varepsilon^2) = \int_0^\infty d\alpha f(\alpha i e^{-i\varepsilon/2})\exp\left(\alpha i e^{-i\varepsilon/2}(p_\varepsilon^2 - m^2)\right). \qquad (39)$$

In (36)

$$x^2(a) \equiv \eta_{\alpha\beta}(a)x^\alpha x^\beta = a^{-1}(x^0)^2 - \mathbf{x}^2, \qquad (40)$$

$$p^2(a) \equiv \eta^{\alpha\beta}(a)p_\alpha p_\beta = a \cdot p_0^2 - \mathbf{p}^2.$$

Let us formulate the exact statement.

**Proposition 2.** *Let the Euclidean correlator of the scalar field in the coordinate and momentum representations be regular and have the form (34), where*

1) *The function $\tilde{D}(z)$ is holomorphic in $C\backslash(-\infty, 0]$, and $D(z)$ is holomorphic in $C\backslash(-\infty, -m^2]$ ;*
2) *For $D(p_E^2)$ the relation (35) is valid, where $f(\alpha)$ satisfies the condition $1^o$ of the Definition 1;*
3) *$\tilde{D}(z) = O(z^\alpha), z \to 0, z \in C\backslash(-\infty, 0]$, where $\alpha > -D/2$ and $D(z) = O(z^T), z \to 0, z \in C\backslash(-\infty, 0]$.*

*Then: a) for all $a \in C\backslash[0, +\infty)$ and $p \in R^D$ relation (37) is valid;*

   *b) for all $a \in C\backslash[0, +\infty)$*

$$\tilde{D}(-x^2(a))\ u\ (-a)^{1/2}F^{-1}\left(D(-p^2(a))\right)(x)$$

*are regular generalized functions from $S'(R^D)$, holomorphic in parameter $a$ in $C\backslash[0, +\infty)$ and connected by the relation*

$$\tilde{D}(-x^2(a)) = (-a)^{1/2}\ F^{-1}\left(D(-p^2(a))\right)(x). \qquad (41)$$

*Moreover, if*

4) $\left(\underline{\varphi(x)\varphi(0)}\right)(a)$ *for all $a \in C\backslash[0, +\infty)$ is a generalized function of $S'(R^D)$, holomorphic with respect to $a$ in $C\backslash[0, +\infty)$ and satisfying the normalization condition*

$$(\underline{\varphi(x^0, \mathbf{x})\varphi(0)})\,(a) = (\underline{\varphi(|a|^{-1/2}x^0, \mathbf{x})\varphi(0)})_E \qquad (42)$$

*for $a < 0$, then*

*c) the relations (36) are valid.*

**Proof.**

a) For $a \in C\setminus[0, +\infty)$ and any fixed $p \in R^D$ the integral in (37) exists and is holomorphic with respect to $a$ in $C\setminus[0, +\infty)$. The left hand side of (37) $\forall\, p \in R^D$ is holomorphic with respect to $a$ in $C\setminus[0, +\infty)$. If $a < 0$, the equality (41) is satisfied. Then (by the uniqueness theorem for holomorphic functions) (37) holds for all $p \in R^D$ and $a \in C\setminus[0, +\infty)$.

b) Regularity of $\tilde{D}(-x^2(a))$ and holomorphic dependence on $a$ in $C\setminus[0, +\infty)$ follow from 3) and inequalities (29). Regularity of $(-a)^{1/2}D(-p^2(a))$ follows from the relation (37), the inequalities (29) and constraints on $f(\alpha)$. Indeed, from the form of $f(\alpha)$ and Abel's theorem [5] we obtain

$$d_2(a)|m^2 - p^2(a)|^{1-s-\delta} \leq \left|(-a)^{1/2}D(-p_2(a))\right|$$
$$\leq d_1(a)|m^2 - p^2(a)|^{1-s+\delta} \qquad (43)$$

where $\delta > 0$ is arbitrary, and $|m^2 - p^2(a)|$ is sufficiently large $[d_1(a) > 0,\ d_2(a) > 0]$, or due to (29)

$$c_2(a)|m^2 + p_E^2|^{1-s-\delta} \leq \left|(-a)^{1/2}D\left(-p^2(a)\right)\right| \leq c_1(a)|m^2 - p_E^2|^{1-s+\delta} \quad (44)$$

where $a \in C\setminus(0, +\infty), \delta > 0, p_E^2 > N\ [c_1(a) > 0,\ c_2(a) > 0]$.

The upper bound in (44) can be made uniform in $\{a\,|\,r < |a| < R, |arg(-a)| < \pi - \gamma\}$ at any $0 < r < R < \infty, 0 < \gamma < \pi$. This readily implies the regularity of $(-a)^{1/2}D(-p^2(a))$ and the holomorphic dependence on $a$ in $C\setminus[0, +\infty)$.

As to the equality (41), at $a = -1$ it is satisfied by definition, in the case of $a < 0$ it may be obtained from the equations at $a = -1$ by a simple dilation:

$$x_0 \to |x_0\,|a|^{-1/2};\quad p_0 \to p_0|a|^{1/2}. \qquad (45)$$

The left and right hand sides of (41) are generalized functions from $S'(R^D)$ holomorphically dependent on parameter $a$ in $C\setminus[0, +\infty)$ (the Fourier transform in $S'(R^D)$ preserves the holomorphic dependence on the parameter). For $a < 0$ the relation (41) is satisfied, then (41) is valid in $C\setminus[0, +\infty)$ by virtue of the proper uniqueness theorem.

c) It is sufficient to prove

$$\left(\underline{\varphi(x)\varphi(0)}\right)(a) = \tilde{D}(-x^2(a)).$$

For $a < 0$ the equality is satisfied by virtue of (41) and (42), and for

$a \in C\setminus[0,+\infty)$ due to: B), the condition 4), the equality for $a < 0$ and the uniqueness theorem for holomorphic in parameter generalized functions from $S'(R^D)$.

**Remark.** The normalization condition for the (42) is natural one. For example, a formally defined $a$-correlator (entire one)

$$\langle\varphi(x)\varphi(0)\rangle(a) = \frac{\int D\varphi\, \varphi(x)\varphi(0) \exp(iS(\varphi,a))}{\int D\varphi \exp(iS(\varphi,a))}, \qquad (46)$$

where

$$S(\varphi,a) = \int (d^D x)(a) \left(\frac{1}{2}(\partial\varphi)^2(a) - V(\varphi)\right)$$

and

$$(d^D x)(a) \equiv d^D x \sqrt{_2 a^{-1}} = d^D x(-a)^{-1/2}/i\,;\ (\partial\varphi)^2(a) = a\cdot(\partial_0\varphi)^2 - (\boldsymbol{\partial}\varphi)^2$$

satisfies the condition (42), as can be easily seen by replacing the integration of variables in (46):

$$\varphi(x^0, \mathbf{x}) = \varphi'(x^0|a|^{-1/2}, \mathbf{x}).$$

It follows from the definition of the proper $\varepsilon$-correlator and Proposition 2, that

$$(\underline{\Phi\Phi})_\varepsilon(p) = P(p,\varepsilon)(-i)e^{i\varepsilon/2}F(m^2 - k_\varepsilon^2) \qquad (47)$$

and

$$F(m^2 - k_\varepsilon^2) = O\left((m^2 + k_E^2)^{-1-s+\delta}\right), \qquad (48)$$

$$(m^2 + k_E^2)^{-1-s-\delta} = O\left(F(m^2 - k_\varepsilon^2)\right)$$

for $k_E^2 \to \infty$. Here $0 < \varepsilon < 2\pi$, and $\delta > 0$ is arbitrary.

Due to (48) it is natural to call the value

$$d = \dim P - 2 - 2s \qquad (49)$$

(where $\dim P$ is the dimension of the polynomial in (32)) as the ultraviolet dimension of $\varepsilon$-correlator (32).

It follows from Proposition 2 that in the simplest case of a scalar field, the regularity of the Euclidean correlator in the coordinate representation guarantees, under some reasonable assumptions, the regularity of the $\varepsilon$-correlator in the coordinate representation ($0 < \varepsilon < 2\pi$). This statement can be made more general.

**Example 4.** In the case of a free scalar massive field in $R^4$ we have

$$D(z) = \frac{1}{m^2+z} \; ; \; \widetilde{D}(z) = \frac{mK_1(mz^{1/2})}{4\pi^2 z^{1/2}},$$

$$\underline{(\varphi(x)\varphi(0))}_\varepsilon = \frac{mK_1(m(-x_\varepsilon^2)^{1/2})}{4\pi^2(-x_\varepsilon^2)^{1/2}}$$

for $0 < \varepsilon < 2\pi, \; m > 0$.

**5. Conclusion.** Thus, the regularization of the pseudo-Euclidean singularities by complex ε-metric, proposed in this article, is free of disadvantages inherent in the covariant regularization. At the same time, in the next article of the author it will be proved that Feynman integrals composed of proper ε-correlators and renormalized in the α-representation determine covariant generalized functions in the limit $\varepsilon = +0$, i.e., when the regularization is removed, the covariance is restored.